# A STUDY OF CYBER SECURITY CHALLENGES AND ITS EMERGNING TRENDS ON LATEST TECHNOLOGIES

## G.NIKHITA REDDY[1] , G.J.UGANDER REDDY[2]

[1]  B.E, CSE second year at Chaitanya Bharathi Institute of Technology, Osmania University, Hyderabad., India
[2]  B.E, M.B.A. and Founder Director - Peridot Technologies, Hyderabad, India

## ABSTRACT

Cyber Security plays an important role in the field of information technology .Securing the information have become one of the biggest challenges in the present day. When ever we think about the cyber security the first thing that comes to our mind is 'cyber crimes' which are increasing immensely day by day. Various Governments and companies are taking many measures in order to prevent these cyber crimes. Besides various measures cyber security is still a very big concern to many. This paper mainly focuses on challenges faced by cyber security on the latest technologies .It also focuses on latest about the cyber security techniques, ethics and the trends changing the face of cyber security.

**Keywords:** cyber security, cyber crime, cyber ethics, social media, cloud computing, android apps.

## 1. INTRODUCTION

Today man is able to send and receive any form of data may be an e-mail or an audio or video just by the click of a button but did he ever think how securely his data id being transmitted or sent to the other person safely without any leakage of information?? The answer lies in cyber security. Today Internet is the fastest growing infrastructure in every day life. In today's technical environment many latest technologies are changing the face of the man kind. But due to these emerging technologies we are unable to safeguard our private information in a very effective way and hence these days cyber crimes are increasing day by day. Today more than 60 percent of total commercial transactions are done online, so this field required a high quality of security for transparent and best transactions. Hence cyber security has become a latest issue. The scope of cyber security is not just limited to securing the information in IT industry but also to various other fields like cyber space etc.

Even the latest technologies like cloud computing, mobile computing, E-commerce, net banking etc also needs high level of security. Since these technologies hold some important information regarding a person their security has become a must thing. Enhancing cyber security and protecting critical information infrastructures are essential to each nation's security and economic wellbeing. Making the Internet safer (and protecting Internet users) has become integral to the development of new services as well as governmental policy. The fight against cyber crime needs a comprehensive and a safer approach. Given that technical measures alone cannot prevent any crime, it is critical that law enforcement agencies are allowed to investigate and prosecute cyber crime effectively. Today many nations and governments are imposing strict laws on cyber securities in order to prevent the loss of some important information. Every individual must also be trained on this cyber security and save themselves from these increasing cyber crimes

## 2. CYBER CRIME

Cyber crime is a term for any illegal activity that uses a computer as its primary means of commission and theft. The U.S. Department of Justice expands the definition of cyber crime to include any illegal activity that uses a computer for the storage of evidence. The growing list of cyber crimes includes crimes that have been made possible by computers, such as network intrusions and the dissemination of computer viruses, as well as computer-based variations of existing crimes, such as identity theft, stalking, bullying and terrorism which have become as major problem to people and nations. Usually in common man's language cyber crime may be defined as crime committed using a computer and the internet to steel a person's identity or sell contraband or stalk victims or disrupt operations with malevolent programs. As day by day technology is playing in major role in a person's life the cyber crimes also will increase along with the technological advances.

## 3. CYBER SECURITY

Privacy and security of the data will always be top security measures that any organization takes care. We are presently living in a world where all the information is maintained in a digital or a cyber form. Social networking sites provide a space where users feel safe as they interact with friends and family. In the case of home users, cyber-criminals would continue to target social media sites to steal personal data. Not only social networking but also during bank transactions a person must take all the required security measures.

| Incidents | Jan-June 2012 | Jan-June 2013 | % Increase/ (decrease) |
|---|---|---|---|
| Fraud | 2439 | 2490 | 2 |
| Intrusion | 2203 | 1726 | (22) |
| Spam | 291 | 614 | 111 |
| Malicious code | 353 | 442 | 25 |
| Cyber Harassment | 173 | 233 | 35 |
| Content related | 10 | 42 | 320 |
| Intrusion Attempts | 55 | 24 | (56) |
| Denial of services | 12 | 10 | (17) |
| Vulnerability reports | 45 | 11 | (76) |
| Total | 5581 | 5592 | |

Table I

The above Comparison of Cyber Security Incidents reported to Cyber999 in Malaysia from January–June 2012 and 2013 clearly exhibits the cyber security threats. As crime is increasing even the security measures are also increasing. According to the survey of U.S. technology and healthcare executives nationwide, Silicon Valley Bank found that companies believe cyber attacks are a serious threat to both their data and their business continuity.

- 98% of companies are maintaining or increasing their cyber security resources and of those, half are increasing resources devoted to online attacks this year
- The majority of companies are preparing for when, not if, cyber attacks occur
- Only one-third are completely confident in the security of their information and even less confident about the security measures of their business partners.

There will be new attacks on Android operating system based devices, but it will not be on massive scale. The fact tablets share the same operating system as smart phones means they will be soon targeted by the same malware as those platforms. The number of malware specimens for Macs would continue to grow, though much less than in the case of PCs. Windows 8 will allow users to develop applications for virtually any device (PCs, tablets and smart phones) running Windows 8, so it will be possible to develop malicious applications like those for Android, hence these are some of the predicted trends in cyber security.

## 4. TRENDS CHANGING CYBER SECURITY

Here mentioned below are some of the trends that are having a huge impact on cyber security.

### 4.1 Web servers:

The threat of attacks on web applications to extract data or to distribute malicious code persists. Cyber criminals distribute their malicious code via legitimate web servers they've compromised. But data-stealing attacks, many of which get the attention of media, are also a big threat. Now, we need a greater emphasis on protecting web servers and web applications. Web servers are especially the best platform for these cyber criminals to steal the data. Hence one must always use a safer browser especially during important transactions in order not to fall as a prey for these crimes.

## 4.2 Cloud computing and its services

These days all small, medium and large companies are slowly adopting cloud services. In other words the world is slowly moving towards the clouds. This latest trend presents a big challenge for cyber security, as traffic can go around traditional points of inspection. Additionally, as the number of applications available in the cloud grows, policy controls for web applications and cloud services will also need to evolve in order to prevent the loss of valuable information. Though cloud services are developing their own models still a lot of issues are being brought up about their security. Cloud may provide immense opportunities but it should always be noted that as the cloud evolves so as its security concerns increase.

## 4.3 APT's and targeted attacks

APT (Advanced Persistent Threat) is a whole new level of cyber crime ware. For years network security capabilities such as web filtering or IPS have played a key part in identifying such targeted attacks (mostly after the initial compromise). As attackers grow bolder and employ more vague techniques, network security must integrate with other security services in order to detect attacks. Hence one must improve our security techniques in order to prevent more threats coming in the future.

## 4.4 Mobile Networks

Today we are able to connect to anyone in any part of the world. But for these mobile networks security is a very big concern. These days firewalls and other security measures are becoming porous as people are using devices such as tablets, phones, PC's etc all of which again require extra securities apart from those present in the applications used. We must always think about the security issues of these mobile networks. Further mobile networks are highly prone to these cyber crimes a lot of care must be taken in case of their security issues.

## 4.5 IPv6: New internet protocol

IPv6 is the new Internet protocol which is replacing IPv4 (the older version), which has been a backbone of our networks in general and the Internet at large. Protecting IPv6 is not just a question of porting IPv4 capabilities. While IPv6 is a wholesale replacement in making more IP addresses available, there are some very fundamental changes to the protocol which need to be considered in security policy. Hence it is always better to switch to IPv6 as soon as possible in order to reduce the risks regarding cyber crime.

## 4.6 Encryption of the code

Encryption is the process of encoding messages (or information) in such a way that eavesdroppers or hackers cannot read it.. In an encryption scheme, the message or information is encrypted using an encryption algorithm, turning it into an unreadable cipher text. This is usually done with the use of an encryption key, which specifies how the message is to be encoded. Encryption at a very beginning level protects data privacy and its integrity. But more use of encryption brings more challenges in cyber security. Encryption is also used to protect data in transit, for example data being transferred via networks (e.g. the Internet, e-commerce), mobile telephones, wireless microphones, wireless intercoms etc. Hence by encrypting the code one can know if there is any leakage of information.

Hence the above are some of the trends changing the face of cyber security in the world. The top network threats are mentioned in below Fig -1.

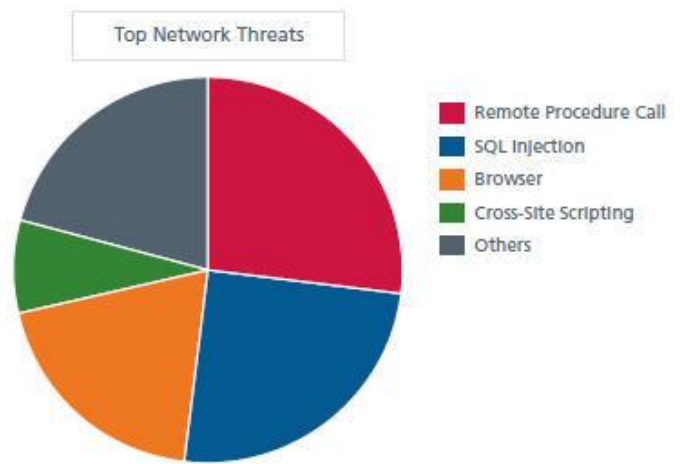

Fig -1

The above pie chart shows about the major threats for networks and cyber security.

## 5. ROLE OF SOCIAL MEDIA IN CYBER SECURITY

As we become more social in an increasingly connected world, companies must find new ways to protect personal information. Social media plays a huge role in cyber security and will contribute a lot to personal cyber threats. Social media adoption among personnel is skyrocketing and so is the threat of attack. Since social media or social networking sites are almost used by most of them every day it has become a huge platform for the cyber criminals for hacking private information and stealing valuable data.

In a world where we're quick to give up our personal information, companies have to ensure they're just as quick in identifying threats, responding in real time, and avoiding a breach of any kind. Since people are easily attracted by these social media the hackers use them as a bait to get the information and the data they require. Hence people must take appropriate measures especially in dealing with social media in order to prevent the loss of their information.

The ability of individuals to share information with an audience of millions is at the heart of the particular challenge that social media presents to businesses. In addition to giving anyone the power to disseminate commercially sensitive information, social media also gives the same power to spread false information, which can be just being as damaging. The rapid spread of false information through social media is among the emerging risks identified in *Global Risks 2013* report.

Though social media can be used for cyber crimes these companies cannot afford to stop using social media as it plays an important role in publicity of a company. Instead, they must have solutions that will notify them of the threat in order to fix it before any real damage is done. However companies should understand this and recognise the importance of analysing the information especially in social conversations and provide appropriate security solutions in order to stay away from risks. One must handle social media by using certain policies and right technologies.

# 6. CYBER SECURITY TECHNIQUES

## 6.1 Access control and password security

The concept of user name and password has been fundamental way of protecting our information. This may be one of the first measures regarding cyber security.

## 6.2 Authentication of data

The documents that we receive must always be authenticated be before downloading that is it should be checked if it has originated from a trusted and a reliable source and that they are not altered. Authenticating of these documents is usually done by the anti virus software present in the devices. Thus a good anti virus software is also essential to protect the devices from viruses.

## 6.3 Malware scanners

This is software that usually scans all the files and documents present in the system for malicious code or harmful viruses. Viruses, worms, and Trojan horses are examples of malicious software that are often grouped together and referred to as malware.

## 6.4 Firewalls

A firewall is a software program or piece of hardware that helps screen out hackers, viruses, and worms that try to reach your computer over the Internet. All messages entering or leaving the internet pass through the firewall present, which examines each message and blocks those that do not meet the specified security criteria. Hence firewalls play an important role in detecting the malware.

## 6.5 Anti-virus software

Antivirus software is a computer program that detects, prevents, and takes action to disarm or remove malicious software programs, such as viruses and worms. Most antivirus programs include an auto-update feature that enables the program to download profiles of new viruses so that it can check for the new viruses as soon as they are discovered. An anti virus software is a must and basic necessity for every system.

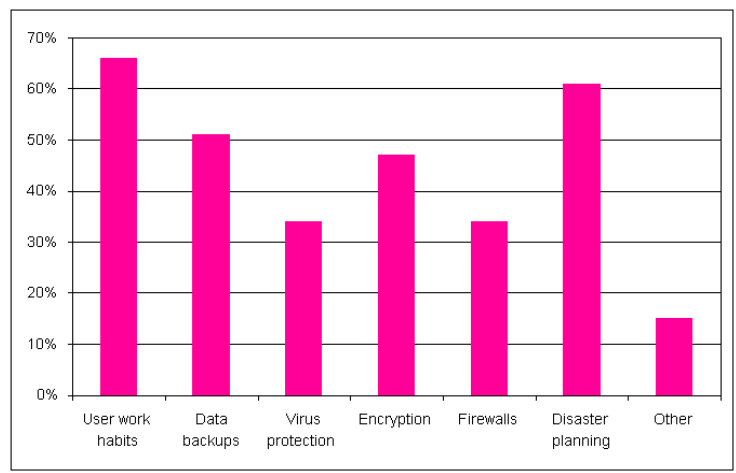

**Table II: Techniques on cyber security**

# 7 CYBER ETHICS

Cyber ethics are nothing but the code of the internet. When we practice these cyber ethics there are good chances of us using the internet in a proper and safer way. The below are a few of them:

- DO use the Internet to communicate and interact with other people. Email and instant messaging make it easy to stay in touch with friends and family members, communicate with work colleagues, and share ideas and information with people across town or halfway around the world
- Don't be a bully on the Internet. Do not call people names, lie about them, send embarrassing pictures of them, or do anything else to try to hurt them.
- Internet is considered as world's largest library with information on any topic in any subject area, so using this information in a correct and legal way is always essential.
- Do not operate others accounts using their passwords.
- Never try to send any kind of malware to other's systems and make them corrupt.
- Never share your personal information to anyone as there is a good chance of others misusing it and finally you would end up in a trouble.
- When you're online never pretend to the other person, and never try to create fake accounts on someone else as it would land you as well as the other person into trouble.
- Always adhere to copyrighted information and download games or videos only if they are permissible.

The above are a few cyber ethics one must follow while using the internet. We are always thought proper rules from out very early stages the same here we apply in cyber space.

# 8. CONCLUSION

Computer security is a vast topic that is becoming more important because the world is becoming highly interconnected, with networks being used to carry out critical transactions. Cyber crime continues to diverge down different paths with each New Year that passes and so does the security of the information. The latest and disruptive technologies, along with the new cyber tools and threats that come to light each day, are challenging organizations with not only how they secure their infrastructure, but how they require new platforms and intelligence to do so. There is no perfect solution for cyber crimes but we should try our level best to minimize them in order to have a safe and secure future in cyber space.